\documentclass{article}
\usepackage{hyperref}
\usepackage{cite}
\usepackage{balance}
\usepackage{comment}
\usepackage{spconf,amsmath,graphicx}
\usepackage{booktabs}
\usepackage{amssymb,amsfonts,amsthm,bm}
\usepackage{mathtools}
\usepackage{tikz}
\usetikzlibrary{patterns}
\usepackage{tikz-cd}
\usepackage{tikz}
\usepackage{tkz-euclide}
\usepackage{pgfplots}
\pgfplotsset{compat=1.17}
\usepackage{tikz-3dplot}
\usetikzlibrary{shapes.geometric}
\usetikzlibrary{calc}
\usetikzlibrary{decorations.markings,arrows.meta,bending}
\usetikzlibrary{math} 
\usetikzlibrary{positioning}
\usetikzlibrary{intersections,through,backgrounds}
\newcommand{\yslant}{0.5}
\newcommand{\xslant}{-0.6}
\tikzset{->-/.style={decoration={markings,
  mark=at position #1 with {\arrow[line width=2pt]{>}}},postaction={decorate}}}

\makeatletter
\def\tikzonnode#1{%
    \pgfutil@ifnextchar[{\tikzonnode@opt#1}{\tikzonnode@opt#1[]}%
}
\def\endtikzonnode{%
    \end{scope}%
}
\def\tikzonnode@opt#1[#2]{%
    \pgfpointanchor{#1}{south west}%
    \pgfgetlastxy\tse@tikz@shift@x\tse@tikz@shift@y
    \begin{scope}[
            shift={(\tse@tikz@shift@x,\tse@tikz@shift@y)},
            x={(#1.south east)},
            y={(#1.north west)},
            #2]%
}

\def\tikzonimage{%
    \pgfutil@ifnextchar[{\tikzonimage@opt}{\tikzonimage@opt[]}%
}
\def\tikzonimage@opt[#1]#2{%
    \begin{tikzpicture}
        \node[inner sep=0] (image) {\includegraphics[#1]{#2}};
        \begin{tikzonnode}{image}%
}
\def\endtikzonimage{%
        \end{tikzonnode}
    \end{tikzpicture}%
}
\makeatother

\theoremstyle{remark}

\title{Causal Linear Topological Filters over a 2-Simplex}
\name{Georg Essl\thanks{This work was made possible by the support of the John Simon Guggenheim Foundation Fellowship.}}
\address{Department of Mathematical Sciences, University of Wisconsin - Milwaukee\\ \url{essl@uwm.edu}}
\tikzset{
   pics/triru/.style args={scale #1}{
        code={
    \begin{scope}[scale=#1]
    \node[bullet={below:a}] (a) at (0:3) {};
    \node[bullet={above left:b}] (b) at (120:3) {};
    \node[bullet={below left:c}] (c) at (240:3) {};
    \node (ba) at ($(b)!0.5!(a)$) {};
    \node (ca) at ($(c)!0.5!(a)$) {};
    \node (bc) at ($(b)!0.5!(c)$) {};
    \draw [fill=gray, fill opacity=0.15,rotate=0]  (a.center) -- (b.center) -- (c.center) -- cycle;
    
    \draw [black,line width=1.5pt,->-/.list={1/2}] 
    (b.center) -- (a.center);
    \draw [black,line width=1.5pt,->-/.list={1/2}] 
    (c.center) -- (a.center);
    \draw [black,line width=1.5pt,->-/.list={1/2}] 
    (b.center) -- (c.center);
    
    \draw[black,line width=1.5pt,-{Latex[bend]}] (-60:0.5) arc(-60:240:0.5);
    \end{scope}
   }
   }
}

\tikzset{
   pics/trif/.style args={scale #1}{
        code={
    \begin{scope}[scale=#1]
    \node (a) at (0:3) {};
    \node (b) at (120:3) {};
    \node (c) at (240:3) {};
    \node[] (ba) at ($(b)!0.5!(a)$) {};
    \node[] (ca) at ($(c)!0.5!(a)$) {};
    \node[] (bc) at ($(b)!0.5!(c)$) {};
    \node (abc) at (0,0) {cba};
    \draw [fill=gray, line width=1.5pt,opacity=0.15,rotate=0]  (a.center) -- (b.center) -- (c.center) -- cycle;
    \draw [black,line width=2pt,->] (ba.center) -- (abc);
    \draw [black,line width=2pt,->] (bc.center) -- (abc);

    \coordinate (A2) at ($(a)!0.5cm!-90:(b)$);
    \coordinate (B2) at ($(b)!0.5cm!90:(a)$);
    \coordinate (A21) at ($(A2)!0.5cm!-90:(B2)$);
    \coordinate (B21) at ($(B2)!0.5cm!90:(A2)$);
    \node[label={[shift={(0.45,-0.10)}]ba}] (BA2) at ($(B2)!0.5!(A2)$) {};
    \coordinate (AB21) at ($(BA2)!0.5cm!90:(A2)$);
    \coordinate (AB22) at ($(BA2)!0.5cm!-90:(B2)$);
    \draw [black,shorten >=.2cm,line width=2pt,->] (A21) -- (AB21);
    \draw [black,shorten >=.2cm,line width=2pt,->] (B21) -- (AB22);
    \draw [name path=ABp,line width=1.5pt] (B2) -- (A2);
    \draw [black,line width=2pt,->] (BA2.center) -- (abc);

    \coordinate (A3) at ($(a)!-0.5cm!-90:(c)$);
    \coordinate (C3) at ($(c)!-0.5cm!90:(a)$);
    \coordinate (A31) at ($(A3)!0.5cm!90:(C3)$);
    \coordinate (C31) at ($(C3)!0.5cm!-90:(A3)$);
    \node[label={[shift={(0.45,-0.6)}]ca}] (CA2) at ($(A3)!0.5!(C3)$) {};
    \coordinate (AC31) at ($(CA2)!0.5cm!-90:(A3)$);
    \coordinate (AC32) at ($(CA2)!0.5cm!90:(C3)$);
    \draw [black,shorten >=.2cm,line width=2pt,->] (A31) -- (AC31);
    \draw [black,shorten >=.2cm,line width=2pt,->] (C31) -- (AC32);
    \draw [name path=ACp,line width=1.5pt] (A3) -- (C3);
    \draw [black,line width=2pt,->] (CA2.center) -- (abc);

    \coordinate (B4) at ($(b)!0.5cm!-90:(c)$);
    \coordinate (C4) at ($(c)!0.5cm!90:(b)$);
    \coordinate (B41) at ($(B4)!0.5cm!-90:(C4)$);
    \coordinate (C41) at ($(C4)!0.5cm!90:(B4)$);
    \node[label={[shift={(-0.05,-0.4)}]cb}] (CB2) at ($(C4)!0.5!(B4)$) {};
    \coordinate (CB41) at ($(CB2)!0.5cm!90:(B4)$);
    \coordinate (CB42) at ($(CB2)!0.5cm!-90:(C4)$);
    \draw [black,shorten >=.2cm,line width=2pt,->] (B41) -- (CB41);
    \draw [black,shorten >=.2cm,line width=2pt,->] (C41) -- (CB42);
    \draw [line width=1.5pt] (B4) -- (C4);
    \draw [black,line width=2pt,->] (CB2.center) -- (abc);

    \path (A2) -- (B2) coordinate[pos=-2](AA2) coordinate[pos=2](BB2);
    \path (A3) -- (C3) coordinate[pos=-2](AA3) coordinate[pos=2](CC3);
    \path (B4) -- (C4) coordinate[pos=-2](BB4) coordinate[pos=2](CC4);
    \path [name path=AA2BB2] (AA2)--(BB2);
    \path [name path=AA3CC3] (AA3)--(CC3);
    \path [name path=BB4CC4] (BB4)--(CC4);
    \path [name intersections={of=AA2BB2 and AA3CC3,by=INTA}];
    \fill [black] (INTA) circle (8pt);
    \node [label=-90:a] at (INTA) {};
    \path [name intersections={of=AA3CC3 and BB4CC4,by=INTA}];
    \fill [black] (INTA) circle (8pt);
    \node [label=-90:c] at (INTA) {};
    \path [name intersections={of=BB4CC4 and AA2BB2,by=INTA}];
    \fill [black] (INTA) circle (8pt);
    \node [label=-90:b] at (INTA) {};

    \end{scope}
   }
   }
}

\tikzset{
   pics/trifnl/.style args={scale #1}{
        code={
    \begin{scope}[scale=#1]
    \node (a) at (0:3) {};
    \node (b) at (120:3) {};
    \node (c) at (240:3) {};
    \node[] (ba) at ($(b)!0.5!(a)$) {};
    \node[] (ca) at ($(c)!0.5!(a)$) {};
    \node[] (bc) at ($(b)!0.5!(c)$) {};
    \node[draw=blue!50,text=black,line width=1.5pt,circle] (abc) at (0,0) {$\mathbb{R}^{2n}$};
    \draw [fill=gray, line width=1.5pt,opacity=0.15,rotate=0]  (a.center) -- (b.center) -- (c.center) -- cycle;

    \coordinate (A2) at ($(a)!0.5cm!-90:(b)$);
    \coordinate (B2) at ($(b)!0.5cm!90:(a)$);
    \coordinate (A21) at ($(A2)!0.5cm!-90:(B2)$);
    \coordinate (B21) at ($(B2)!0.5cm!90:(A2)$);
    \node[label={[shift={(0.15,-0.1)}]$\mathbb{R}^{n}$}] (BA2) at ($(B2)!0.5!(A2)$) {};
    \coordinate (AB21) at ($(BA2)!0.5cm!90:(A2)$);
    \coordinate (AB22) at ($(BA2)!0.5cm!-90:(B2)$);
    \draw [black,shorten >=.2cm,line width=1pt,->] (A21) -- (AB21) node[midway,above right] {$r_x$};
    \draw [black,shorten >=.2cm,line width=1pt,->] (B21) -- (AB22) node[midway,above right] {$s_x$};
    \draw [name path=ABp,line width=1.5pt] (B2) -- (A2);
    \draw [black,line width=1pt,->] (BA2.center) -- (abc) node[midway,left] {$r_x$};

    \coordinate (A3) at ($(a)!-0.5cm!-90:(c)$);
    \coordinate (C3) at ($(c)!-0.5cm!90:(a)$);
    \coordinate (A31) at ($(A3)!0.5cm!90:(C3)$);
    \coordinate (C31) at ($(C3)!0.5cm!-90:(A3)$);
    \node[label={[shift={(0.15,-0.6)}]$\mathbb{R}^{n}$}] (CA2) at ($(A3)!0.5!(C3)$) {};
    \coordinate (AC31) at ($(CA2)!0.5cm!-90:(A3)$);
    \coordinate (AC32) at ($(CA2)!0.5cm!90:(C3)$);
    \draw [black,shorten >=.2cm,line width=1pt,->] (A31) -- (AC31) node[midway,below right] {$r_y$};
    \draw [black,shorten >=.2cm,line width=1pt,->] (C31) -- (AC32) node[midway,below right] {$s_y$};
    \draw [name path=ACp,line width=1.5pt] (A3) -- (C3);
    \draw [black,line width=1pt,->] (CA2.center) -- (abc) node[midway,left] {$r_y$};

    \coordinate (B4) at ($(b)!0.5cm!-90:(c)$);
    \coordinate (C4) at ($(c)!0.5cm!90:(b)$);
    \coordinate (B41) at ($(B4)!0.5cm!-90:(C4)$);
    \coordinate (C41) at ($(C4)!0.5cm!90:(B4)$);
    \node[] (CB2) at ($(C4)!0.5!(B4)$) {};
    \node[draw=blue!50,text=black,line width=1.5pt,circle] at ([shift={(-0.99,0)}]CB2) {$\mathbb{R}^{2n\! +\! 2}$};
    \coordinate (CB41) at ($(CB2)!0.5cm!90:(B4)$);
    \coordinate (CB42) at ($(CB2)!0.5cm!-90:(C4)$);
    \draw [black,shorten >=.2cm,line width=1pt,->] (B41) -- (CB41) node[midway,label={[shift={(-0.32,-0.15)}]:{$r_{xi}$}}] {};
    \draw [black,shorten >=.2cm,line width=1pt,->] (C41) -- (CB42) node[midway,label={[shift={(-0.32,-0.7)}]:{$r_{yi}$}}] {};
    \draw [line width=1.5pt] (B4) -- (C4);
    \draw [black,line width=1pt,->] (CB2.center) -- (abc) node[midway,label={[shift={(0.1,-0.9)}]:{$s_{xy}$}}] {}; 

    \path (A2) -- (B2) coordinate[pos=-0.1675](AA2) coordinate[pos=1.1675](BB2);
    \path (A3) -- (C3) coordinate[pos=-0.1675](AA3) coordinate[pos=1.1675](CC3);
    \path (B4) -- (C4) coordinate[pos=-0.1675](BB4) coordinate[pos=1.1675](CC4);
    \path [name path=AA2BB2] (AA2)--(BB2);
    \path [name path=AA3CC3] (AA3)--(CC3);
    \path [name path=BB4CC4] (BB4)--(CC4);
    \path [name intersections={of=AA2BB2 and AA3CC3,by=INTA}];
    \fill [black] (INTA) circle (6pt);
    \coordinate[] (INTAA) at ($(INTA)+(6,0)$);
    \draw [black,shorten >=.4cm,shorten <=.4cm,line width=1.5pt] (INTA) --  (INTAA);
    \fill [black] (INTAA) circle (6pt);
    \node[] (INTAM) at ($(INTA)!0.5!(INTAA)$) {};
    \draw [green,line width=1.5pt] (INTAM) ellipse (3.5cm and 0.8cm);
    \coordinate (INTAM1) at ($(INTA)!0.5cm!-90:(INTAM)$);
    \coordinate (INTAM2) at ($(INTAA)!0.5cm!90:(INTAM)$);
    \node[] (INTAM3) at ($(INTAM1)!0.5!(INTAM2)$) {$\mathbb{R}^n$};
    \draw [black,shorten <=.4cm,line width=1pt,->] (INTAM1) -- (INTAM3) node[midway,label={[shift={(0.3,-0.7)}]:{$m$}}] {};
    \draw [black,shorten <=.4cm,line width=1pt,->] (INTAM2) -- (INTAM3) node[midway,label={[shift={(-0.2,-0.7)}]:{$r$}}] {};
    \node [] at ([shift={(0.15,-0.8)}]INTAA) {$\mathbb{R}^{n+1}$};

    \node [draw=red!50,text=black,line width=1.5pt,circle] at ([shift={(0.25,-0.8)}]INTA) {$\mathbb{R}^{2n}$};
    \path [name intersections={of=AA3CC3 and BB4CC4,by=INTAI}];
    \fill [black] (INTAI) circle (6pt);
    \node [label=below:$\mathbb{R}^{n\! +\! 1}$] at (INTAI) {};
    \path [name intersections={of=BB4CC4 and AA2BB2,by=INTA}];
    \fill [black] (INTA) circle (6pt);
    \node [label=above:$\mathbb{R}^{n\! +\! 1}$] at (INTA) {};
  
    \node [label=right:\bf{State} $\mathcal{S}_s$] (LABEL) at (CA2 |- INTAI) {}; 
    \end{scope}
   }
   }
}

\tikzset{
   pics/trifnli/.style args={scale #1}{
        code={
    \begin{scope}[scale=#1]
    \node (a) at (0:3) {};
    \node (b) at (120:3) {};
    \node (c) at (240:3) {};
    \node[] (ba) at ($(b)!0.5!(a)$) {};
    \node[] (ca) at ($(c)!0.5!(a)$) {};
    \node[] (bc) at ($(b)!0.5!(c)$) {};
    \node[draw=blue!50,text=black,line width=1.5pt,circle] (abc) at (0,0) {$0$};
    \draw [fill=gray, line width=1.5pt,opacity=0.15,rotate=0]  (a.center) -- (b.center) -- (c.center) -- cycle;

    \coordinate (A2) at ($(a)!0.5cm!-90:(b)$);
    \coordinate (B2) at ($(b)!0.5cm!90:(a)$);
    \coordinate (A21) at ($(A2)!0.5cm!-90:(B2)$);
    \coordinate (B21) at ($(B2)!0.5cm!90:(A2)$);
    \node[label={[shift={(0.15,-0.1)}]$0$}] (BA2) at ($(B2)!0.5!(A2)$) {};
    \coordinate (AB21) at ($(BA2)!0.5cm!90:(A2)$);
    \coordinate (AB22) at ($(BA2)!0.5cm!-90:(B2)$);
    \draw [black,shorten >=.2cm,line width=1pt,->] (A21) -- (AB21);
    \draw [black,shorten >=.2cm,line width=1pt,->] (B21) -- (AB22);
    \draw [name path=ABp,line width=1.5pt] (B2) -- (A2);
    \draw [black,line width=1pt,->] (BA2.center) -- (abc);

    \coordinate (A3) at ($(a)!-0.5cm!-90:(c)$);
    \coordinate (C3) at ($(c)!-0.5cm!90:(a)$);
    \coordinate (A31) at ($(A3)!0.5cm!90:(C3)$);
    \coordinate (C31) at ($(C3)!0.5cm!-90:(A3)$);
    \node[label={[shift={(0.15,-0.6)}]$0$}] (CA2) at ($(A3)!0.5!(C3)$) {};
    \coordinate (AC31) at ($(CA2)!0.5cm!-90:(A3)$);
    \coordinate (AC32) at ($(CA2)!0.5cm!90:(C3)$);
    \draw [black,shorten >=.2cm,line width=1pt,->] (A31) -- (AC31);
    \draw [black,shorten >=.2cm,line width=1pt,->] (C31) -- (AC32);
    \draw [name path=ACp,line width=1.5pt] (A3) -- (C3);
    \draw [black,line width=1pt,->] (CA2.center) -- (abc);

    \coordinate (B4) at ($(b)!0.5cm!-90:(c)$);
    \coordinate (C4) at ($(c)!0.5cm!90:(b)$);
    \coordinate (B41) at ($(B4)!0.5cm!-90:(C4)$);
    \coordinate (C41) at ($(C4)!0.5cm!90:(B4)$);
    \node[] (CB2) at ($(C4)!0.5!(B4)$) {};
    \node[draw=blue!50,text=black,line width=1.5pt,circle] at ([shift={(-0.99,0)}]CB2) {$0$};
    \coordinate (CB41) at ($(CB2)!0.5cm!90:(B4)$);
    \coordinate (CB42) at ($(CB2)!0.5cm!-90:(C4)$);
    \draw [black,shorten >=.2cm,line width=1pt,->] (B41) -- (CB41);
    \draw [black,shorten >=.2cm,line width=1pt,->] (C41) -- (CB42);
    \draw [line width=1.5pt] (B4) -- (C4);
    \draw [black,line width=1pt,->] (CB2.center) -- (abc);

    \path (A2) -- (B2) coordinate[pos=-0.1675](AA2) coordinate[pos=1.1675](BB2);
    \path (A3) -- (C3) coordinate[pos=-0.1675](AA3) coordinate[pos=1.1675](CC3);
    \path (B4) -- (C4) coordinate[pos=-0.1675](BB4) coordinate[pos=1.1675](CC4);
    \path [name path=AA2BB2] (AA2)--(BB2);
    \path [name path=AA3CC3] (AA3)--(CC3);
    \path [name path=BB4CC4] (BB4)--(CC4);
    \path [name intersections={of=AA2BB2 and AA3CC3,by=INTA}];
    \fill [black] (INTA) circle (8pt);
    \coordinate[] (INTAA) at ($(INTA)+(6,0)$);
    \draw [black,shorten >=.4cm,shorten <=.4cm,line width=1.5pt] (INTA) --  (INTAA);
    \fill [black] (INTAA) circle (8pt);
    \node[] (INTAM) at ($(INTA)!0.5!(INTAA)$) {};
    \draw [green,line width=1.5pt] (INTAM) ellipse (3.5cm and 1cm);
    \coordinate (INTAM1) at ($(INTA)!0.5cm!-90:(INTAM)$);
    \coordinate (INTAM2) at ($(INTAA)!0.5cm!90:(INTAM)$);
    \node[] (INTAM3) at ($(INTAM1)!0.5!(INTAM2)$) {$0$};
    \draw [black,shorten <=.4cm,line width=1pt,->] (INTAM1) -- (INTAM3);
    \draw [black,shorten <=.4cm,line width=1pt,->] (INTAM2) -- (INTAM3);
    \node [] at ([shift={(0.15,-0.8)}]INTAA) {$\mathbb{R}$};

    \node [draw=red!50,text=black,line width=1.5pt,circle] at ([shift={(0.25,-0.8)}]INTA) {$0$};
    \path [name intersections={of=AA3CC3 and BB4CC4,by=INTAI}];
    \fill [black] (INTAI) circle (8pt);
    \node [label=below:$\mathbb{R}$] at (INTAI) {};
    \path [name intersections={of=BB4CC4 and AA2BB2,by=INTA}];
    \fill [black] (INTA) circle (8pt);
    \node [label=above:$\mathbb{R}$] at (INTA) {};
  
    \node [label=right:\bf{Input} $\mathcal{S}_i$] (LABEL) at (CA2 |- INTAI) {}; 
    \end{scope}
   }
   }
}

\tikzset{
   pics/trifnlo/.style args={scale #1}{
        code={
    \begin{scope}[scale=#1]
    \node (a) at (0:3) {};
    \node (b) at (120:3) {};
    \node (c) at (240:3) {};
    \node[] (ba) at ($(b)!0.5!(a)$) {};
    \node[] (ca) at ($(c)!0.5!(a)$) {};
    \node[] (bc) at ($(b)!0.5!(c)$) {};
    \node[draw=blue!50,text=black,line width=1.5pt,circle] (abc) at (0,0) {$\mathbb{R}$};
    \draw [fill=gray, line width=1.5pt,opacity=0.15,rotate=0]  (a.center) -- (b.center) -- (c.center) -- cycle;

    \coordinate (A2) at ($(a)!0.5cm!-90:(b)$);
    \coordinate (B2) at ($(b)!0.5cm!90:(a)$);
    \coordinate (A21) at ($(A2)!0.5cm!-90:(B2)$);
    \coordinate (B21) at ($(B2)!0.5cm!90:(A2)$);
    \node[label={[shift={(0.15,-0.1)}]$0$}] (BA2) at ($(B2)!0.5!(A2)$) {};
    \coordinate (AB21) at ($(BA2)!0.5cm!90:(A2)$);
    \coordinate (AB22) at ($(BA2)!0.5cm!-90:(B2)$);
    \draw [black,shorten >=.2cm,line width=1pt,->] (A21) -- (AB21);
    \draw [black,shorten >=.2cm,line width=1pt,->] (B21) -- (AB22);
    \draw [name path=ABp,line width=1.5pt] (B2) -- (A2);
    \draw [black,line width=1pt,->] (BA2.center) -- (abc);

    \coordinate (A3) at ($(a)!-0.5cm!-90:(c)$);
    \coordinate (C3) at ($(c)!-0.5cm!90:(a)$);
    \coordinate (A31) at ($(A3)!0.5cm!90:(C3)$);
    \coordinate (C31) at ($(C3)!0.5cm!-90:(A3)$);
    \node[label={[shift={(0.15,-0.6)}]$0$}] (CA2) at ($(A3)!0.5!(C3)$) {};
    \coordinate (AC31) at ($(CA2)!0.5cm!-90:(A3)$);
    \coordinate (AC32) at ($(CA2)!0.5cm!90:(C3)$);
    \draw [black,shorten >=.2cm,line width=1pt,->] (A31) -- (AC31);
    \draw [black,shorten >=.2cm,line width=1pt,->] (C31) -- (AC32);
    \draw [name path=ACp,line width=1.5pt] (A3) -- (C3);
    \draw [black,line width=1pt,->] (CA2.center) -- (abc);

    \coordinate (B4) at ($(b)!0.5cm!-90:(c)$);
    \coordinate (C4) at ($(c)!0.5cm!90:(b)$);
    \coordinate (B41) at ($(B4)!0.5cm!-90:(C4)$);
    \coordinate (C41) at ($(C4)!0.5cm!90:(B4)$);
    \node[] (CB2) at ($(C4)!0.5!(B4)$) {};
    \node[draw=blue!50,text=black,line width=1.5pt,circle] at ([shift={(-0.99,0)}]CB2) {$0$};
    \coordinate (CB41) at ($(CB2)!0.5cm!90:(B4)$);
    \coordinate (CB42) at ($(CB2)!0.5cm!-90:(C4)$);
    \draw [black,shorten >=.2cm,line width=1pt,->] (B41) -- (CB41);
    \draw [black,shorten >=.2cm,line width=1pt,->] (C41) -- (CB42);
    \draw [line width=1.5pt] (B4) -- (C4);
    \draw [black,line width=1pt,->] (CB2.center) -- (abc);

    \path (A2) -- (B2) coordinate[pos=-0.1675](AA2) coordinate[pos=1.1675](BB2);
    \path (A3) -- (C3) coordinate[pos=-0.1675](AA3) coordinate[pos=1.1675](CC3);
    \path (B4) -- (C4) coordinate[pos=-0.1675](BB4) coordinate[pos=1.1675](CC4);
    \path [name path=AA2BB2] (AA2)--(BB2);
    \path [name path=AA3CC3] (AA3)--(CC3);
    \path [name path=BB4CC4] (BB4)--(CC4);
    \path [name intersections={of=AA2BB2 and AA3CC3,by=INTA}];
    \fill [black] (INTA) circle (8pt);
    \coordinate[] (INTAA) at ($(INTA)+(6,0)$);
    \draw [black,shorten >=.4cm,shorten <=.4cm,line width=1.5pt] (INTA) --  (INTAA);
    \fill [black] (INTAA) circle (8pt);
    \node[] (INTAM) at ($(INTA)!0.5!(INTAA)$) {};
    \draw [green,line width=1.5pt] (INTAM) ellipse (3.5cm and 1cm);
    \coordinate (INTAM1) at ($(INTA)!0.5cm!-90:(INTAM)$);
    \coordinate (INTAM2) at ($(INTAA)!0.5cm!90:(INTAM)$);
    \node[] (INTAM3) at ($(INTAM1)!0.5!(INTAM2)$) {$0$};
    \draw [black,shorten <=.4cm,line width=1pt,->] (INTAM1) -- (INTAM3);
    \draw [black,shorten <=.4cm,line width=1pt,->] (INTAM2) -- (INTAM3);
    \node [] at ([shift={(0.15,-0.8)}]INTAA) {$\mathbb{R}$};

    \node [draw=red!50,text=black,line width=1.5pt,circle] at ([shift={(0.25,-0.8)}]INTA) {$0$};
    \path [name intersections={of=AA3CC3 and BB4CC4,by=INTAI}];
    \fill [black] (INTAI) circle (8pt);
    \node [label=below:$\mathbb{R}$] at (INTAI) {};
    \path [name intersections={of=BB4CC4 and AA2BB2,by=INTA}];
    \fill [black] (INTA) circle (8pt);
    \node [label=above:$\mathbb{R}$] at (INTA) {};
  
    \node [label=right:\bf{Output} $\mathcal{S}_o$] (LABEL) at (CA2 |- INTAI) {}; 
    \end{scope}
   }
   }
}

\tikzset{
   pics/tribnl/.style args={scale #1}{
        code={
    \begin{scope}[scale=#1]
    \node (a) at (180:3) {};
    \node (b) at (300:3) {};
    \node (c) at (60:3) {};
    \node[] (ba) at ($(b)!0.5!(a)$) {};
    \node[] (ca) at ($(c)!0.5!(a)$) {};
    \node[] (bc) at ($(b)!0.5!(c)$) {};
    \node[draw=blue!50,text=black,line width=1.5pt,circle] (abc) at (0,0) {$\mathbb{R}^{2n}$};
    \draw [fill=gray, line width=1.5pt,opacity=0.15,rotate=0]  (a.center) -- (b.center) -- (c.center) -- cycle;

    \coordinate (A2) at ($(a)!0.5cm!-90:(b)$);
    \coordinate (B2) at ($(b)!0.5cm!90:(a)$);
    \coordinate (A21) at ($(A2)!0.5cm!-90:(B2)$);
    \coordinate (B21) at ($(B2)!0.5cm!90:(A2)$);
    \node[label={[shift={(-0.15,-0.6)}]$\mathbb{R}^{n}$}] (BA2) at ($(B2)!0.5!(A2)$) {};
    \coordinate (AB21) at ($(BA2)!0.5cm!90:(A2)$);
    \coordinate (AB22) at ($(BA2)!0.5cm!-90:(B2)$);
    \draw [black,shorten >=.2cm,line width=1pt,->] (A21) -- (AB21) node[midway,below left] {$s_y$};
    \draw [black,shorten >=.2cm,line width=1pt,->] (B21) -- (AB22) node[midway,below left] {$r$};
    \draw [name path=ABp,line width=1.5pt] (B2) -- (A2);
    \draw [black,line width=1pt,->] (BA2.center) -- (abc) node[midway,above left] {$r_y$};

    \coordinate (A3) at ($(a)!-0.5cm!-90:(c)$);
    \coordinate (C3) at ($(c)!-0.5cm!90:(a)$);
    \coordinate (A31) at ($(A3)!0.5cm!90:(C3)$);
    \coordinate (C31) at ($(C3)!0.5cm!-90:(A3)$);
    \node[label={[shift={(-0.15,-0.1)}]$\mathbb{R}^{n}$}] (CA2) at ($(A3)!0.5!(C3)$) {};
    \coordinate (AC31) at ($(CA2)!0.5cm!-90:(A3)$);
    \coordinate (AC32) at ($(CA2)!0.5cm!90:(C3)$);
    \draw [black,shorten >=.2cm,line width=1pt,->] (A31) -- (AC31) node[midway,above left] {$s_x$};
    \draw [black,shorten >=.2cm,line width=1pt,->] (C31) -- (AC32) node[midway,above left] {$r$};
    \draw [name path=ACp,line width=1.5pt] (A3) -- (C3);
    \draw [black,line width=1pt,->] (CA2.center) -- (abc) node[midway,below left] {$r_x$};

    \coordinate (B4) at ($(b)!0.5cm!-90:(c)$);
    \coordinate (C4) at ($(c)!0.5cm!90:(b)$);
    \coordinate (B41) at ($(B4)!0.5cm!-90:(C4)$);
    \coordinate (C41) at ($(C4)!0.5cm!90:(B4)$);
    \node[] (CB2) at ($(C4)!0.5!(B4)$) {};
    \node[draw=blue!50,text=black,line width=1.5pt,circle] at ([shift={(0.99,0)}]CB2) {$\mathbb{R}^{2n\! +\! 2}$};
    \coordinate (CB41) at ($(CB2)!0.5cm!90:(B4)$);
    \coordinate (CB42) at ($(CB2)!0.5cm!-90:(C4)$);
    \draw [black,shorten >=.2cm,line width=1pt,->] (B41) -- (CB41) node[midway,label={[shift={(0.3,-0.7)}]:{$r_{yi}$}}] {};
    \draw [black,shorten >=.2cm,line width=1pt,->] (C41) -- (CB42) node[midway,label={[shift={(0.3,-0.2)}]:{$r_{xi}$}}] {};
    \draw [line width=1.5pt] (B4) -- (C4);
    \draw [black,line width=1pt,->] (CB2.center) -- (abc) node[midway,label={[shift={(-0.1,-0.9)}]:{$r_{xy}$}}] {};

    \path (A2) -- (B2) coordinate[pos=-0.1675](AA2) coordinate[pos=1.1675](BB2);
    \path (A3) -- (C3) coordinate[pos=-0.1675](AA3) coordinate[pos=1.1675](CC3);
    \path (B4) -- (C4) coordinate[pos=-0.1675](BB4) coordinate[pos=1.1675](CC4);
    \path [name path=AA2BB2] (AA2)--(BB2);
    \path [name path=AA3CC3] (AA3)--(CC3);
    \path [name path=BB4CC4] (BB4)--(CC4);
    \path [name intersections={of=AA2BB2 and AA3CC3,by=INTAI}];
    \fill [black] (INTAI) circle (6pt);
    \node [label={[shift={(-0.25,-0.8)}]$\mathbb{R}^{n\! +\! 1}$}] at (INTAI) {};
    \path [name intersections={of=AA3CC3 and BB4CC4,by=INTA}];
    \fill [black] (INTA) circle (6pt);
    \node [label=above:$\mathbb{R}^{n\! +\! 1}$] at (INTA) {};
    \path [name intersections={of=BB4CC4 and AA2BB2,by=INTA}];
    \fill [black] (INTA) circle (6pt);
    \node [label=below:$\mathbb{R}^{n\! +\! 1}$] at (INTA) {};
    \node (LABEL) at (INTAI |- INTA) {{\bf State} $\mathcal{S}_s$};
    \end{scope}
   }
   }
}

\tikzset{
   pics/tribnli/.style args={scale #1}{
        code={
    \begin{scope}[scale=#1]
    \node (a) at (180:3) {};
    \node (b) at (300:3) {};
    \node (c) at (60:3) {};
    \node[] (ba) at ($(b)!0.5!(a)$) {};
    \node[] (ca) at ($(c)!0.5!(a)$) {};
    \node[] (bc) at ($(b)!0.5!(c)$) {};
    \node[draw=blue!50,text=black,line width=1.5pt,circle] (abc) at (0,0) {$0$};
    \draw [fill=gray, line width=1.5pt,opacity=0.15,rotate=0]  (a.center) -- (b.center) -- (c.center) -- cycle;

    \coordinate (A2) at ($(a)!0.5cm!-90:(b)$);
    \coordinate (B2) at ($(b)!0.5cm!90:(a)$);
    \coordinate (A21) at ($(A2)!0.5cm!-90:(B2)$);
    \coordinate (B21) at ($(B2)!0.5cm!90:(A2)$);
    \node[label={[shift={(-0.15,-0.6)}]$0$}] (BA2) at ($(B2)!0.5!(A2)$) {};
    \coordinate (AB21) at ($(BA2)!0.5cm!90:(A2)$);
    \coordinate (AB22) at ($(BA2)!0.5cm!-90:(B2)$);
    \draw [black,shorten >=.2cm,line width=1pt,->] (A21) -- (AB21);
    \draw [black,shorten >=.2cm,line width=1pt,->] (B21) -- (AB22);
    \draw [name path=ABp,line width=1.5pt] (B2) -- (A2);
    \draw [black,line width=1.0pt,->] (BA2.center) -- (abc);

    \coordinate (A3) at ($(a)!-0.5cm!-90:(c)$);
    \coordinate (C3) at ($(c)!-0.5cm!90:(a)$);
    \coordinate (A31) at ($(A3)!0.5cm!90:(C3)$);
    \coordinate (C31) at ($(C3)!0.5cm!-90:(A3)$);
    \node[label={[shift={(-0.15,-0.1)}]$0$}] (CA2) at ($(A3)!0.5!(C3)$) {};
    \coordinate (AC31) at ($(CA2)!0.5cm!-90:(A3)$);
    \coordinate (AC32) at ($(CA2)!0.5cm!90:(C3)$);
    \draw [black,shorten >=.2cm,line width=1pt,->] (A31) -- (AC31);
    \draw [black,shorten >=.2cm,line width=1pt,->] (C31) -- (AC32);
    \draw [name path=ACp,line width=1.5pt] (A3) -- (C3);
    \draw [black,line width=1pt,->] (CA2.center) -- (abc);

    \coordinate (B4) at ($(b)!0.5cm!-90:(c)$);
    \coordinate (C4) at ($(c)!0.5cm!90:(b)$);
    \coordinate (B41) at ($(B4)!0.5cm!-90:(C4)$);
    \coordinate (C41) at ($(C4)!0.5cm!90:(B4)$);
    \node[] (CB2) at ($(C4)!0.5!(B4)$) {};
    \node[draw=blue!50,text=black,line width=1.5pt,circle] at ([shift={(0.99,0)}]CB2) {$0$};
    \coordinate (CB41) at ($(CB2)!0.5cm!90:(B4)$);
    \coordinate (CB42) at ($(CB2)!0.5cm!-90:(C4)$);
    \draw [black,shorten >=.2cm,line width=1pt,->] (B41) -- (CB41);
    \draw [black,shorten >=.2cm,line width=1pt,->] (C41) -- (CB42);
    \draw [line width=1.5pt] (B4) -- (C4);
    \draw [black,line width=1pt,->] (CB2.center) -- (abc);

    \path (A2) -- (B2) coordinate[pos=-0.1675](AA2) coordinate[pos=1.1675](BB2);
    \path (A3) -- (C3) coordinate[pos=-0.1675](AA3) coordinate[pos=1.1675](CC3);
    \path (B4) -- (C4) coordinate[pos=-0.1675](BB4) coordinate[pos=1.1675](CC4);
    \path [name path=AA2BB2] (AA2)--(BB2);
    \path [name path=AA3CC3] (AA3)--(CC3);
    \path [name path=BB4CC4] (BB4)--(CC4);
    \path [name intersections={of=AA2BB2 and AA3CC3,by=INTAI}];
    \fill [black] (INTAI) circle (8pt);
    \node [label={[shift={(-0.25,-0.8)}]$\mathbb{R}$}] at (INTAI) {};
    \path [name intersections={of=AA3CC3 and BB4CC4,by=INTA}];
    \fill [black] (INTA) circle (8pt);
    \node [label=above:$\mathbb{R}$] at (INTA) {};
    \path [name intersections={of=BB4CC4 and AA2BB2,by=INTA}];
    \fill [black] (INTA) circle (8pt);
    \node [label=below:$\mathbb{R}$] at (INTA) {};
    \node [align = right] (LABEL) at (INTAI |- INTA) {{\bf Input} $\mathcal{S}_i$};
    \end{scope}
   }
   }
}

\tikzset{
   pics/tribnlo/.style args={scale #1}{
        code={
    \begin{scope}[scale=#1]
    \node (a) at (180:3) {};
    \node (b) at (300:3) {};
    \node (c) at (60:3) {};
    \node[] (ba) at ($(b)!0.5!(a)$) {};
    \node[] (ca) at ($(c)!0.5!(a)$) {};
    \node[] (bc) at ($(b)!0.5!(c)$) {};
    \node[draw=blue!50,text=black,line width=1.5pt,circle] (abc) at (0,0) {$\mathbb{R}$};
    \draw [fill=gray, line width=1.5pt,opacity=0.15,rotate=0]  (a.center) -- (b.center) -- (c.center) -- cycle;

    \coordinate (A2) at ($(a)!0.5cm!-90:(b)$);
    \coordinate (B2) at ($(b)!0.5cm!90:(a)$);
    \coordinate (A21) at ($(A2)!0.5cm!-90:(B2)$);
    \coordinate (B21) at ($(B2)!0.5cm!90:(A2)$);
    \node[label={[shift={(-0.15,-0.6)}]$0$}] (BA2) at ($(B2)!0.5!(A2)$) {};
    \coordinate (AB21) at ($(BA2)!0.5cm!90:(A2)$);
    \coordinate (AB22) at ($(BA2)!0.5cm!-90:(B2)$);
    \draw [black,shorten >=.2cm,line width=1pt,->] (A21) -- (AB21);
    \draw [black,shorten >=.2cm,line width=1pt,->] (B21) -- (AB22);
    \draw [name path=ABp,line width=1.5pt] (B2) -- (A2);
    \draw [black,line width=1pt,->] (BA2.center) -- (abc);

    \coordinate (A3) at ($(a)!-0.5cm!-90:(c)$);
    \coordinate (C3) at ($(c)!-0.5cm!90:(a)$);
    \coordinate (A31) at ($(A3)!0.5cm!90:(C3)$);
    \coordinate (C31) at ($(C3)!0.5cm!-90:(A3)$);
    \node[label={[shift={(-0.15,-0.1)}]$0$}] (CA2) at ($(A3)!0.5!(C3)$) {};
    \coordinate (AC31) at ($(CA2)!0.5cm!-90:(A3)$);
    \coordinate (AC32) at ($(CA2)!0.5cm!90:(C3)$);
    \draw [black,shorten >=.2cm,line width=1pt,->] (A31) -- (AC31);
    \draw [black,shorten >=.2cm,line width=1pt,->] (C31) -- (AC32);
    \draw [name path=ACp,line width=1.5pt] (A3) -- (C3);
    \draw [black,line width=1pt,->] (CA2.center) -- (abc);

    \coordinate (B4) at ($(b)!0.5cm!-90:(c)$);
    \coordinate (C4) at ($(c)!0.5cm!90:(b)$);
    \coordinate (B41) at ($(B4)!0.5cm!-90:(C4)$);
    \coordinate (C41) at ($(C4)!0.5cm!90:(B4)$);
    \node[] (CB2) at ($(C4)!0.5!(B4)$) {};
    \node[draw=blue!50,text=black,line width=1.5pt,circle] at ([shift={(0.99,0)}]CB2) {$0$};
    \coordinate (CB41) at ($(CB2)!0.5cm!90:(B4)$);
    \coordinate (CB42) at ($(CB2)!0.5cm!-90:(C4)$);
    \draw [black,shorten >=.2cm,line width=1pt,->] (B41) -- (CB41);
    \draw [black,shorten >=.2cm,line width=1pt,->] (C41) -- (CB42);
    \draw [line width=1.5pt] (B4) -- (C4);
    \draw [black,line width=1pt,->] (CB2.center) -- (abc);

    \path (A2) -- (B2) coordinate[pos=-0.1675](AA2) coordinate[pos=1.1675](BB2);
    \path (A3) -- (C3) coordinate[pos=-0.1675](AA3) coordinate[pos=1.1675](CC3);
    \path (B4) -- (C4) coordinate[pos=-0.1675](BB4) coordinate[pos=1.1675](CC4);
    \path [name path=AA2BB2] (AA2)--(BB2);
    \path [name path=AA3CC3] (AA3)--(CC3);
    \path [name path=BB4CC4] (BB4)--(CC4);
    \path [name intersections={of=AA2BB2 and AA3CC3,by=INTAI}];
    \fill [black] (INTAI) circle (8pt);
    \node [label={[shift={(-0.25,-0.8)}]$\mathbb{R}$}] at (INTAI) {};
    \path [name intersections={of=AA3CC3 and BB4CC4,by=INTA}];
    \fill [black] (INTA) circle (8pt);
    \node [label=above:$\mathbb{R}$] at (INTA) {};
    \path [name intersections={of=BB4CC4 and AA2BB2,by=INTA}];
    \fill [black] (INTA) circle (8pt);
    \node [label=below:$\mathbb{R}$] at (INTA) {};
    \node [align = right] (LABEL) at (INTAI |- INTA) {{\bf Output} $\mathcal{S}_o$};
    \end{scope}
   }
   }
}

\tikzset{
   pics/tris/.style args={scale #1}{
        code={
    \begin{scope}[scale=#1]
    \node (a) at (0:3) {$\mathbb{S}$};
    \node (b) at (120:3) {};
    \node (c) at (240:3) {};
    \node[] (ba) at ($(b)!0.5!(a)$) {};
    \node[] (ca) at ($(c)!0.5!(a)$) {};
    \node[] (bc) at ($(b)!0.5!(c)$) {};
    \node (abc) at (0,0) {cba};
    \draw [fill=gray, line width=1.5pt,opacity=0.15,rotate=0]  (a.center) -- (b.center) -- (c.center) -- cycle;
    \draw [black,line width=2pt,->] (ba.center) -- (abc);
    \draw [black,line width=2pt,->] (bc.center) -- (abc);

    \coordinate (A2) at ($(a)!0.5cm!-90:(b)$);
    \coordinate (B2) at ($(b)!0.5cm!90:(a)$);
    \node[label={[shift={(0.45,-0.10)}]ba}] (BA2) at ($(B2)!0.5!(A2)$) {};
    \draw [line width=1.5pt] (B2) -- (A2);
    \draw [black,line width=2pt,->] (BA2.center) -- (abc);

    \coordinate (A3) at ($(a)!-0.5cm!-90:(c)$);
    \coordinate (C3) at ($(c)!-0.5cm!90:(a)$);
    \node[label={[shift={(0.45,-0.6)}]ca}] (CA2) at ($(A3)!0.5!(C3)$) {};
    \draw [line width=1.5pt] (A3) -- (C3);
    \draw [black,line width=2pt,->] (CA2.center) -- (abc);

    \coordinate (B4) at ($(b)!0.5cm!-90:(c)$);
    \coordinate (C4) at ($(c)!0.5cm!90:(b)$);
    \node[label={[shift={(-0.05,-0.4)}]cb}] (CB2) at ($(C4)!0.5!(B4)$) {};
    \draw [line width=1.5pt] (B4) -- (C4);
    \draw [black,line width=2pt,->] (CB2.center) -- (abc);

    \end{scope}
   }
   }
}

\tikzset{
   pics/trird/.style args={scale #1}{
        code={
    \begin{scope}[scale=#1]
    \node[bullet={below:a}] (a) at (0:3) {};
    \node[bullet={above left:b}] (b) at (120:3) {};
    \node[bullet={below left:c}] (c) at (240:3) {};
    \draw [fill=gray, fill opacity=0.15,rotate=0]  (a.center) -- (b.center) -- (c.center) -- cycle;
    
    \draw [black,line width=1.5pt,->-/.list={1/2}] 
    (b.center) -- (a.center);
    \draw [black,line width=1.5pt,->-/.list={1/2}] 
    (c.center) -- (a.center);
    \draw [black,line width=1.5pt,->-/.list={1/2}] 
    (c.center) -- (b.center);
    
    \draw[black,line width=1.5pt,-{Latex[bend]}] (-120:0.5) arc(240:-60:0.5);
    \end{scope}
   }
   }
}

\tikzset{
   pics/trild/.style args={scale #1}{
        code={
    \begin{scope}[scale=#1]
    \node[bullet={below:a}] (a) at (180:3) {};
    \node[bullet={below right:c}] (c) at (300:3) {};
    \node[bullet={above right:b}] (b) at (60:3) {};
    \draw [fill=gray, fill opacity=0.15,rotate=0]  (a.center) -- (b.center) -- (c.center) -- cycle;
    
    \draw [black,line width=1.5pt,->-/.list={1/2}] 
    (a.center) -- (b.center);
    \draw [black,line width=1.5pt,->-/.list={1/2}] 
    (a.center) -- (c.center);
    \draw [black,line width=1.5pt,->-/.list={1/2}] 
    (b.center) -- (c.center);
    
    \draw[black,line width=1.5pt,-{Latex[bend]}] (-120:0.5) arc(240:-60:0.5);
    \end{scope}
   }
    }
}

\tikzset{
   pics/trilu/.style args={scale #1}{
        code={
    \begin{scope}[scale=#1]
    \node[bullet={below:a}] (a) at (180:3) {};
    \node[bullet={below right:c}] (c) at (300:3) {};
    \node[bullet={above right:b}] (b) at (60:3) {};
    \draw [fill=gray, fill opacity=0.15,rotate=0]  (a.center) -- (b.center) -- (c.center) -- cycle;
    
    \draw [black,line width=1.5pt,->-/.list={1/2}] 
    (a.center) -- (b.center);
    \draw [black,line width=1.5pt,->-/.list={1/2}] 
    (a.center) -- (c.center);
    \draw [black,line width=1.5pt,->-/.list={1/2}] 
    (c.center) -- (b.center);
    
    \draw[black,line width=1.5pt,-{Latex[bend]}] (-60:0.5) arc(-60:240:0.5);
    \end{scope}
   }
    }
}

\tikzset{
   pics/lirl/.style args={scale #1}{
        code={
    \begin{scope}[scale=#1]
        \node[bullet={below:}] (a) at (0,0) {};
        \node[bullet={below:}] (b) at (6,0) {};
        \node[] (c) at (9,0) {};
        \draw[black,line width=1.5pt,->-/.list={1/2}] (a.center) -- (b.center);
    \end{scope}
    }
    }
}
\begin{document}
%
\maketitle
\begin{abstract}
Topological filters via sheaves generalize the classical linear translation-invariant filter theory by attaching the filter computation locally to a simplicial topological space. This paper develops topological filters for causal signal flow over a 2-simplex. Our construction retains the established construction for 1-simplices and we show how an apparent conflict in the sheaf assignment can be resolved by a concurrent extension, which introduces an auxiliary 1-simplex that computes the resolution. Furthermore, we discuss how singularities formed by double cone connections can be resolved.
\end{abstract}
\section{Introduction}
\label{sec:intro}
Topological signal processing via sheaves is an approach that associates local data and computation with a computationally friendly topological space, such as a simplicial complex. The general methodology, which can be used for both linear and nonlinear applications was introduced by Robinson \cite{robinson2014topological}. In this paper we are only concerned with the linear theory, and in particular the extension of classical linear translation-invariant (LTI) digital filters to the sheaf-theoretic case. Classical single-input-single-output (SISO) linear translation-invariant digital filter theory can be completely embedded in the theory of sheaf-theoretic topological filters \cite{robinson2014topological,robinson2013understanding,Essl2020TopIIR}.

While in principle the underlying algebraic topology generalizes to higher dimensions via constructions over n-dimensional simplices with dimension greater than one, explicit constructions are rare \cite[see p.41]{robinson2014topological} and \cite{robinson2017sheaf} and a general theory that allows the connection of the existing topological filter with a higher dimensional simplex and develop sheaf map assignments is still missing. The goal of this paper is to provide an explicit construction for the 2-simplex as the topological space over which a topological filter is given that reflects linear translation-invariant properties.

In this work we view topological filters over a 2-simplex as a modular building block, meaning that the need for global computations is avoided in all our constructions. In particular we are interested in retaining seamless interconnectivity properties with the 1-simplex case, as well as embedding the 1-simplex theory in the 2-simplex construction.

Furthermore we limit our discussion to the case of what we will call {\em causal flow}. This idea captures a notion of traversal over the topological space that reflects the causality of computation. Sheaf maps are expected to be computed in a particular order if they are causally related. Additionally we introduce a notion of {\em concurrency}. Concurrent sheaf maps are not causal, meaning they can be computed in any order but must all be computed before the next causal update is computed.

As an illustrative example, consider the notion of the propagation of a wave front in the plane. The connected component of the wavefront propagates concurrently, while the flow of the wavefront propagates causally.
%
Hence it might be helpful intuition for the reader to consider casual flow to be "time-like" and concurrency to be "space-like", though these associations may not hold for all applications.
%
%
%
%
\section{Related Work}
\label{sec:prior}
The work is a continuation of efforts to expand our understanding of topological signal processing using sheaves \cite{robinson2014topological,Essl2020TopIIR} which in turn is related to work on sheaf cohomology \cite{robinson2014topological}, a spectral theory via sheaf Laplacians and combinatorial Hodge theory over sheaves \cite{hansen2019toward} with applications in graph dynamics such as opinion networks \cite{hansen2021opinion}.
Another related effort is the generalizaton of graph signal processing \cite{Ortega2018GraphSignalProcessing} to simplicial topologies \cite{schaub2021signal2,schaub2021signal,ji2020signal,Barbarossa2020TSPSimplicialComplexes} which in turn is related to studying multi-way interactions \cite{battiston2020networks,Barbarossa2020Multiway}. Lifting construction for traversals of random walks over simplicial complexes proposed in \cite{schaub2020random} provide an alternative approach to oriented traversal using expanded linear space. 
%
%
Graph and simplicial signal processing can be viewed as related to developments in areas of applied and computational topology \cite{edelsbrunner2010computational,ghrist2014elementary,carlsson2009topology} in general and their consideration with respect to time-series signals via persistent homology \cite{perea2015sliding,perea2019topological} in particular. Algebraic and combinatorical topology provide the underlying mathematical theory of these fields. The following texts serve as good introductions with a helpful inclination towards applicable computations \cite{giblin2013graphs,munkres2000topology}.
%
%
%
\section{Simplicial Topology}
\label{sec:simplicial}
\begin{figure}[t]
\begin{minipage}{0.2\columnwidth}
\centering
\begin{tikzpicture}[baseline={(0,0)},=0.8,bullet/.style={circle,fill,inner
sep=3pt,label={[font=\bfseries]#1}},>=latex]
    \draw [black,line width=1.5pt] 
    (0,0) node[bullet={below left:a}] (a) {};
\end{tikzpicture}
\end{minipage}%
\begin{minipage}{0.40\columnwidth}
\centering
\begin{tikzpicture}[baseline={(0,0)},scale=0.8,bullet/.style={circle,fill,inner
sep=3pt,label={[font=\bfseries]#1}},>=latex,scale=0.40]
    \draw [black,line width=1.5pt,->-/.list={1/1.6}] 
    (-2.598,0) node[bullet={below left:a}] (a) {}
    -- (2.598,0) node[bullet={below right:b}] (b) {};
\end{tikzpicture}
\end{minipage}%
\begin{minipage}{0.40\columnwidth}
\centering
\begin{tikzpicture}[baseline={(0,0)},scale=0.8,bullet/.style={circle,fill,inner
sep=3pt,label={[font=\bfseries]#1}},>=latex,scale=0.40]
    \draw [fill=gray, fill opacity=0.15] 
    (210:3) -- (-30:3) -- (90:3) -- cycle;
    \draw [black,line width=1.5pt,->-/.list={1/6,1/2,5/6}] 
    (210:3) node[bullet={below left:a}] (a) {}
    -- (-30:3) node[bullet={below right:b}] (b) {}
    -- (90:3) node[bullet={above:c}] (c) {}
    -- cycle;
    \draw[line width=1.5pt,-{Latex[bend]}] (-60:0.5) arc(-60:240:0.5);
\end{tikzpicture}
\end{minipage}\par\smallskip
\begin{minipage}{0.2\columnwidth}
\center
\color{black}
\bf 0-simplex\\
\end{minipage}%
\begin{minipage}{0.40\columnwidth}
\center
\color{black}
\bf 1-simplex\\
\end{minipage}%
\begin{minipage}{0.40\columnwidth}
\center
\color{black}
\bf 2-simplex\\
\end{minipage}\\
\begin{minipage}{0.2\columnwidth}
\center
\color{black}
\bf $\{a\}$
\end{minipage}%
\begin{minipage}{0.40\columnwidth}
\center
\color{black}
\bf $\{a,b,$\\$\{a,b\}\}$
\end{minipage}%
\begin{minipage}{0.40\columnwidth}
\center
\color{black}
\bf $\{a,b,c,$\\$\{a,b\},\{b,c\},\{c,a\},$\\$\{a,b,c\}\}$
\end{minipage}%
\caption{Low-dimensional simplices and their combinatorial description as simplicial sets.}\label{fig:simplices}
\end{figure}
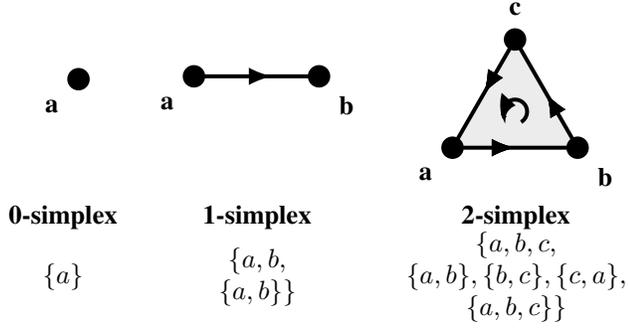

Simplices between dimension zero and two are topological versions of points, lines, and areas, though only the connectivity of these concepts is topological. This means that we can define computationally-friendly abstract oriented simplices as the combinatorics and list manipulation of sets. It can be observed from Figure \ref{fig:simplices} that removal of an entry of a higher-order simplex set is equivalent to one of the lower order sets. For example, the deletion of $b$ from ${a,b,c}$ in a 2-simplex is the same as ${a,c}$ which is precisely the 1-simplex opposite the removed 0-simplex $b$. For this reason a removal operation from the list is equivalent to a face map and can be used as its definition \cite{eilenberg1950semi,friedman2012elementary}. 
For shorthand we will call $n$-simplices that are faces of $(n+1)$-simplices simply $n$-faces. We call $n$ the dimension of an $n$-simplex. A simplicial complex is the set of simplicies that can be connected by identification at shared faces. A line complex is a simplicial constituted of 1-simplicies identified at shared 0-simplices to form a line-like simplicial complex.
\subsection{Causal Traversal and Concurrency}
We consider the case of a causal traversial over simplices within a simplicial complex of order up to $2$-simplices. This is a generalization of the traversal over 0- and 1-simplices \cite{Essl2020TopIIR}. Figs. \ref{fig:face01layer} and \ref{fig:face10layer} help illustrate the cases under consideration. In general figures will treat the horizontal direction as causal (unless indicated otherwise) and the vertical direction as concurrent (without exception). The constructions given do retain orientations.



%
\subsection{Sheaves over simplices}
Sheaves are the name for a construction that attaches local data and associated computation to a topological space \cite{robinson2014topological}. In our case the topological space is characterized by simplices, and the maps provided are face maps. Hence a sheaf attaches data to each simplex called a {\em stalk}. Further a sheaf requires that each map over the topology be associated with a sheaf map. In our simplicial topologies the maps are the face maps, and hence we construct suitable sheaf maps over each face map. Finally there is a requirement that local data is consistent through compositions of sheaf maps (for a detailed exposition see \cite{robinson2014topological}).
\section{Topological Filter Sheaf}

Robinson defined a set of sheaf maps that constitute a {\em topological filter} \cite{robinson2014topological}:
\begin{align}\label{eq:topfilt}
    \mathcal{S}_i\xleftarrow{i}\mathcal{S}_{s}\xrightarrow{o}\mathcal{S}_o
\end{align}
This structure is attached to each simplex, though potentially with zero input or output, hence constitutes as a whole a stalk. In principle the data that constitutes this sheaf construction is flexible and is not confined to linear maps or vector spaces \cite{robinson2014topological,Essl2021topologize}. However, in this work we will confine ourselves to sheaves over vector spaces acted upon by linear maps representing linear translation-invariant digital filters \cite{Essl2020TopIIR}.
%

%
%
%
%
%
%
%
%
\subsection{Causal LTI Filters over a 1-simplex}
\begin{figure}[t]
\centering
\begin{tikzcd}
\text{input layer} \arrow{r} & 0 &\arrow{l} \mathbb{R} \arrow{r} & 0 & \arrow{l}  \\
\text{state layer} \arrow{r} & \mathbb{R}^{n} \arrow{u} \arrow{d}  &\arrow{l}{r} \mathbb{R}^{n+1} \arrow{u}{i} \arrow{d}{o} \arrow{r}{s} & \mathbb{R}^{n} \arrow{u} \arrow{d} & \arrow{l}   \\
\text{output layer}\arrow{r} & 0 &\arrow{l} \mathbb{R} \arrow{r} & 0 & \arrow{l}   
\end{tikzcd}
\caption{Vector spaces and sheaf maps of a topological filter over a 1-simplex. The vertical structure corresponds to the topological filter sheaf. The horizontal structure captures input, state, and output layers.\label{fig:sheafallpole}}
\end{figure}
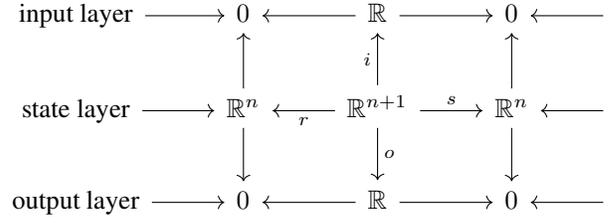
\noindent The sheaf structure over a 1-simplex for general linear time invariant filters can be shown to be as depicted in Fig. \ref{fig:sheafallpole}. The vertical structure are each topological filters as defined by (\ref{eq:topfilt}) and the causal direction is left to right. It can be shown \cite{Essl2020TopIIR} that the linear maps in this structure for the general IIR filter are as follows, with coefficients $a_\bullet$ and $b_\bullet$ corresponding to the usual filter coefficients:
%
%
\begin{align}
\begin{split}
s: (x_0,x_1,...,x_{n-1},x) \rightarrow& (x_1,x_2,...,x_{n-1},\\&  x+\sum_{j=1}^{n}{-a_j\cdot x_{n-j}})
\end{split}\label{eq:iirfb}\\
r: (x_0,x_1,...,x_{n-1},x) \rightarrow& (x_0,x_1,...,x_{n-1})\\
i: (x_0,x_1,...,x_{n-1},x) \rightarrow& (x)\\
o: (x_0,x_1,...,x_{n-1},x) \rightarrow& (b_0 \cdot x + \sum_{i=1}^n{b_i\cdot x_{n-i}})\label{eq:iirff}
\end{align}
%
%



\section{Causal LTI filter over a 2-simplex}

\begin{figure}[t]
\centering
\begin{tikzpicture}[baseline={(0,0)},scale=0.8,bullet/.style={circle,fill,inner
sep=3pt,label={[font=\bfseries]#1}},>=latex]
\pic(name) at (0,3) {tribnl={scale 0.5}};
\pic(name) at (5,5.0) {tribnli={scale 0.3}};
\pic(name) at (5,1.0) {tribnlo={scale 0.3}};
\end{tikzpicture}
\vspace{-12pt}
    \caption{The sheaf maps between stalks over a 2-simplex as a 0-to-1-signal flow in the input, state, and output sheaf layer.}
    \label{fig:face01layer}
\end{figure}
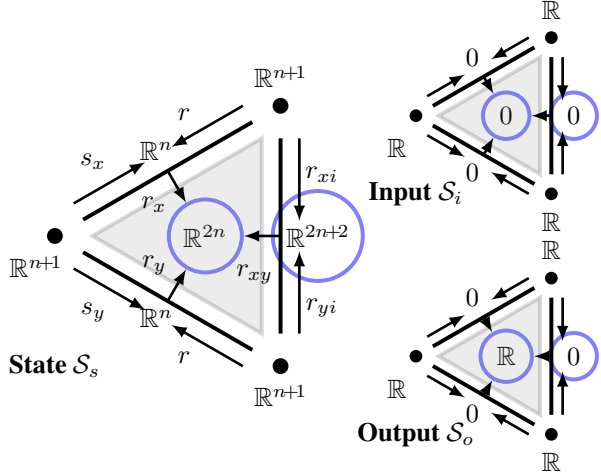



Next we consider the construction of the sheaf maps over a causal topological filter over a 2-simplex. The definition of stalks and the sheaf maps connecting them are described with respect to three layers, which are, from top to bottom, the input layer, the state layer and the output layer. The vertical maps are omitted, though they can usually be read from Fig. \ref{fig:sheafallpole}, which captures the topological filter of eq. \ref{eq:topfilt} of causal 1-faces.



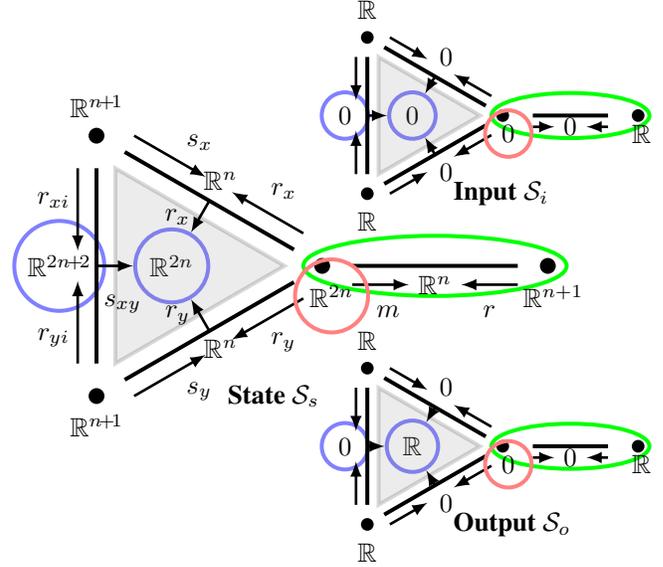
\begin{figure}[t]
\centering
\begin{tikzpicture}[baseline={(0,0)},scale=0.8,bullet/.style={circle,fill,inner
sep=3pt,label={[font=\bfseries]#1}},>=latex]
\pic(name) at (0,3) {trifnl={scale 0.5}};
\pic(name) at (4,5.5) {trifnli={scale 0.3}};
\pic(name) at (4,0) {trifnlo={scale 0.3}};
\end{tikzpicture}
\vspace{-12pt}
    \caption{The sheaf maps between stalks over a 2-simplex as a 1-to-0-signal flow in the input, state, and output sheaf layer.}
    \label{fig:face10layer}
\end{figure}

We will depict the causal direction of traversal as going from left to right. We have two basic configurations of causal traversals over 2-simplices to consider. 
First is a 2-simplex whose causal input connects through a single 0-simplex on the left and provides concurrent 0- and 1-simplices on the right. We will call traversals through a 0-simplex a {\em 0-signal}, and traversals through a 1-simplex with two 0-faces, a {\em 1-signal}.
Hence we will refer to this case as a {\em 0-to-1-signal flow} (Fig. \ref{fig:face01layer}). Second, a 2-simplex whose input is a 1-signal over a concurrent 1-simplex on the left and a 0-signal on the right (Fig. \ref{fig:face10layer}), constituting a {\em 1-to-0-signal flow}. 

The general approach is shared between both cases. First, we retain the topological filter structure of Fig. \ref{fig:sheafallpole} for all 1-faces that are causal. 
They correspond to the sheaf from Fig. \ref{fig:sheafallpole} and the corresponding linear maps of equations (\ref{eq:iirfb})-(\ref{eq:iirff}). This means that all stalks over 0-simplices are already fixed. However, we observe that there is a conflict between consistency requirements and this construction with the rightmost 0-simplex in Fig. \ref{fig:face10layer}, circled in red.

What is further left is to construct the sheaf maps and stalks over the 1-simplex that is concurrent and the internal sheaf maps and the stalk associated with the 2-simplex. The unassigned maps are circled in blue in Figs. \ref{fig:face01layer} and \ref{fig:face10layer}. The resolution of the conflict by {\em concurrent 1-simplex extension} is depicted in green.

\subsection{Concurrent 1-simplex Extension}

Merging two causal 1-simplices at the right 0-simplex in a 1-to-0 signal flow causes a conflict (Fig. \ref{fig:face10layer}, red). By consistency we require that retrieval maps entering into the 0-simplex be injective. Further they cannot be assumed to be identical. Hence the dimension of the vector space is necessarily at least $\mathbb{R}^{2n}$. However, we want to connect to topological filters over 1-simplices which are specified for states of size $\mathbb{R}^{n+1}$ following Fig. \ref{fig:sheafallpole}. We propose a technique we call concurrent 1-simplex extension to resolve the issue. The technique consists of inserting a 1-simplex after the rightmost 0-simplex, and define a merge map $m$ over it (Fig. \ref{fig:face10layer}, green). This simplex is considered concurrent, hence it is computed before any further traversal. Observe that on the right side of the extension we recover the desired 0-signal interface.

The merge map $m: \mathbb{R}^{2n}\rightarrow\mathbb{R}^n$ computes a linear combination between two states, which in turn contains the case of one-parameter linear interpolation of two vectors.
\begin{align}
     m:(x_0,x_1,\cdots,x_{n-1},y_0,y_1,\cdots,y_{n-1})\xrightarrow{}\\
    (a_0\cdot x_0 + b_0\cdot y_0, \cdots a_{n-1}\cdot x_{n-1}+b_{n-1}\cdot y_{n-1}) \label{eq:merging}
\end{align}

\subsection{Consistency Constructions}

All remaining maps are constructed from consistency requirements. The consistency requirement states that all individual compositions of face maps that end on the same simplex need to agree. Furthermore, we want to propagate the state over our filter, hence we are confined to not lose state dimensions. Hence it is natural to compose accumulations of states and inputs as direct sums and their separation as the associated projections. Hence we arrive at the stalks of the locations circled in blue in Figs. \ref{fig:face01layer} and \ref{fig:face10layer}. Furthermore, in order to allow the connection of our 2-simplex with incoming and outgoing 1-simplices, we impose maps to 0 over the concurrent 1-simplex circled blue. This in turn fixed all layers over this simplex.

\subsection{2-simplex Output}

Finally we observe that given all stalks and maps fixed so far, the output over the 2-simplex itself (the blue circle in the middle of the 2-simplex) is not restricted as it is surrounded by maps to 0 under all sheaf maps in the output layer. This gives us the freedom to define an output map $o_2: \mathbb{R}^{2n}\rightarrow\mathbb{R}$. Given that the direct sum of both states from the causal 1-simplex are present, this output map can compute a joint output as linear map from both of these states.

\begin{align}
    o_2:(x_0,x_1,\cdots,x_{n-1},y_0,y_1,\cdots,y_{n-1})\xrightarrow{}\\
    (\sum_{i=0}^{n-1}{a_i\cdot x_{i}} + \sum_{i=0}^{n-1}{b_i\cdot y_{i}}) \label{eq:mixingoutput}
\end{align}


\subsection{Combined State Update Map}

The map $s_{xy}$ is a direct sum of shift maps $s_x$ and $s_y$. The maps $s_x$ and $s_y$ are structurally identical to the state update map $s$ from equation (\ref{eq:iirfb}) with potentially differing coefficients.

\begin{align}
\begin{split}
    s_{xy}: &s_x\oplus s_y =\\
    (&x_0,x_1,...,x_{n-1},x,y_0,y_1,...,y_{n-1},y) \rightarrow \\ (&x_1,x_2,...,x_{n-1},x+\sum_{j=1}^{N}{-a_j\cdot x_{n-j}},\\
    &y_1,y_2,...,y_{n-1},y+\sum_{k=1}^{N}{-b_k\cdot y_{n-k}})
\end{split}
\end{align}

\subsection{Projection maps}

All remaining maps are projections (or alternatively direct sums of injections if read in the inverse direction). We generally use label maps $r$ if they are projections. Subscripts $\Box_x$ and $\Box_y$ refer to the branch on the 2-simplex the map is associated with, or the combination of both branches, and the subscript $\Box_i$ indicates that an input has already been injected into the state sheaf and is hence retained for consistency under the map. Their derivation is straightfoward. They are:

\begin{align}
  \phantom{r_{xy}:}&\begin{alignedat}{1}
    \mathllap{r_x:}& (x_0,x_1,...,x_{n-1},y_0, y_1,...,y_{n-1}) \rightarrow \\
        & (x_0,x_1,...,x_{n-1})
\end{alignedat}    \\
&\begin{alignedat}{1}
    \mathllap{r_y:}& (x_0,x_1,...,x_{n-1},y_0, y_1,...,y_{n-1}) \rightarrow \\
    & (y_0,y_1,...,y_{n-1})
\end{alignedat}    \\
&\begin{alignedat}{1}
    \mathllap{r_{xy}}:& (x_0,x_1,...,x_{n-1},x,y_0, y_1,...,y_{n-1},y)\rightarrow \\ 
    & (x_0,x_1,...,x_{n-1},y_0, y_1,...,y_{n-1})
\end{alignedat}    \\
&\begin{alignedat}{1}
    \mathllap{r_{xi}}:& (x_0,x_1,...,x_{n-1},x,y_0, y_1,...,y_{n-1},y) \rightarrow \\
    & (x_0,x_1,...,x_{n-1},x)
\end{alignedat}    \\
&\begin{alignedat}{1}
    \mathllap{r_{yi}}:& (x_0,x_1,...,x_{n-1},x,y_0, y_1,...,y_{n-1},y) \rightarrow \\
    & (y_0,y_1,...,y_{n-1},y)\\
\end{alignedat}    
\end{align}

\section{Resolution of Singularities of two-sided Cones}

The theory presented here can be extended to treat singularities of a 1-signal flows pinching onto a 0-simplex and then expanding again. Consider a 1-0-signal flow connected to a 0-1-signal flow, forming a two-sided cone. In our given construction, the concurrent 1-simplex extension provides the interconnectivity, but has the effect of collapsing a 1-signal in $\mathbb{R}^{2n}$ onto $\mathbb{R}^n$. While the collapse of the topology onto a 0-simplex is by construction, the collapse of the data above it is optional, and this insight allows us to construct a two-sided cone that does not generate a singularity. The interconnecting state sheaf over the 0-simplex is modified to be $\mathbb{R}^{2n+2}$ hence in the dimension of a 1-signal. If the continuation of the filter map through the identification at this point is linear translation invariant, the solution falls into linear time invariant theory. Specifically maps $s_x$ and $s_y$ swap roles after the singularity hence there is an induced change in orientation. An example of this case is the collapse of a traveling wave front through a local point singularity. The induced change of orientation as the wave front passes through the singularity is well known as wave front reversal \cite[p. 47]{arnold1994topological}.

\section{Conclusions}

In this paper we developed the causal flow of signals over a 2-simplex in the context of sheaf-theoretic topological filters over simplicial topologies. The resulting linear maps allow the connection of individual signals over a 0-simplex to signals over 1-simplicies and vice-versa. For reasons of space the theory of 1-to-1 signal flow is only hinted at through the treatment of the point singularity and will follow in a later publication. Treatment of higher-dimensional simplicial elements such 3-simplices still remain open. This is future work. More broadly the connection of this theory to sheaf-theoretic Hodge theory and sheaf-theoretic Laplacians still needs to be developed.

\vfill\pagebreak


\balance
\bibliographystyle{IEEEtran}
\bibliography{strings,refs}

\end{document}